# Who Are the Phishers? Phishing Scam Detection on Ethereum via Network Embedding

Jiajing Wu, *Senior Member, IEEE*, Qi Yuan, Dan Lin, Wei You, Weili Chen, Chuan Chen, *Member, IEEE*, and Zibin Zheng, *Senior Member, IEEE*

*Abstract*—Recently, blockchain technology has become a topic in the spotlight but also a hotbed of various cybercrimes. Among them, phishing scams on blockchain have been found to make a notable amount of money, thus emerging as a serious threat to the trading security of the blockchain ecosystem. In order to create a favorable environment for investment, an effective method for detecting phishing scams is urgently needed in the blockchain ecosystem. To this end, this article proposes an approach to detect phishing scams on Ethereum by mining its transaction records. Specifically, we first crawl the labeled phishing addresses from two authorized websites and reconstruct the transaction network according to the collected transaction records. Then, by taking the transaction amount and timestamp into consideration, we propose a novel network embedding algorithm called *trans2vec* to extract the features of the addresses for subsequent phishing identification. Finally, we adopt the one-class support vector machine (SVM) to classify the nodes into normal and phishing ones. Experimental results demonstrate that the phishing detection method works effectively on Ethereum, and indicate the efficacy of *trans2vec* over existing state-of-the-art algorithms on feature extraction for transaction networks. This work is the *first* investigation on phishing detection on Ethereum via network embedding and provides insights into how features of large-scale transaction networks can be embedded.

## I. Introduction

BLOCKCHAIN is an open and distributed ledger that can record transactions between two parties efficiently, verifiably, and permanently [1]. Recently, blockchain has become a topic in the spotlight and the generalized blockchain technology is expected to bring profound changes in the fields of finance, science and technology, culture, and politics [2]. One of the most important and famous applications of blockchain in economics is a digital asset (or cryptocurrency). The bitcoin project is the first successful large-scale application of blockchain and the first practical implementation of cryptocurrency.

Ethereum is currently the largest blockchain platform that supports smart contracts and the corresponding cryptocurrency *ether* is the second-largest cyptocurrency [3]. However, along with its high-speed development, Ethereum has also become a hotbed of various cybercrimes [4]. Initial coin offering (ICO) is a financing method for the blockchain industry, which refers to financing through the issuance of tokens. However, till now, more than 10% of ICOs released on Ethereum have been reported to be suffer from a variety of scams, including phishing, Ponzi schemes, etc. [5]. According to a report of *Chainalysis*, a provider of investigation and risk management software for virtual currencies, there were 30 287 victims losing $225 million in the first half of 2017 [6], indicating that financial security has become a critical issue in the blockchain ecosystem.

Besides, among various security issues of blockchain digital cryptocurrency, the number of phishing scams accounts for more than 50% of all cybercrimes in Ethereum since 2017 and this kind of scam has become as a main threat to trading security of Ethereum [7]. A typical phishing scam on Ethereum happened when Bee Token, a blockchain-based home-sharing service, planned to launch its ICO on January 31, 2018. Before the official release of the ICO, phishers sent fake emails to would-be investigators of the ICO and promised them an extra bonus for all the contributions within the next 6 h and a double value of the token within the next two months. This phishing scam eventually swindled nearly $1 million in just 25 h. In order to create a favorable environment for investment on the blockchain ecosystem, an effective method for detection and prevention of phishing scams is urgently needed.

In the past decades, with the rise of online business, phishing scams emerges as a main threat to trading security. By disguising as a trustworthy entity, phishers attempt to obtain the users' sensitive information, such as usernames, passwords, and credit card details. The issue of phishing detection has been widely and extensively discussed and a number of antiphishing methods have been proposed. However, compared with traditional scenarios, phishing scams on Ethereum behave very differently in several aspects.

First, as cryptocurrencies instead of cash become the target of phishing scams, the phishers on Ethereum need to cash the ill-gotten cryptocurrencies through exchanges for fiat money. Second, all the transaction records of public blockchain

Manuscript received November 20, 2019; accepted August 10, 2020. This work was supported in part by the Key-Area Research and Development Program of Guangdong Province under Grant 2019B020214006, and in part by the National Natural Science Foundation of China under Grant 61973325 and Grant U1811462. This article was recommended by Associate Editor F. Wang. *(Corresponding authors: Jiajing Wu; Chuan Chen.)*
The authors are with the School of Data and Computer Science, Sun Yat-sen University, Guangzhou 510006, China (e-mail: wujiajing@mail.sysu.edu.cn; chenchuan@mail.sysu.edu.cn).
Color versions of one or more of the figures in this article are available online at http://ieeexplore.ieee.org.
Digital Object Identifier 10.1109/TSMC.2020.3016821

systems such as Ethereum are publicly accessible, providing us a complete data source for mining the transaction manner of different Ethereum users, which may be useful for phishing detection. Third, as most traditional phishing frauds rely on phishing emails and websites to obtain the users' sensitive information, existing methods of phishing detection usually focus on how to detect the emails or websites containing phishing fraud information [8]. However, phishing methods on Ethereum are usually much more diverse than that of traditional phishing scams and phishing information can be spread in a variety of forms.

Therefore, existing detection methods for traditional phishing frauds cannot be directly applied to solve the phishing detection problem on blockchain platforms like Ethereum. Though important and urgent, the problem of anti-phishing on blockchain ecosystem has never been discussed in current work.

As openness and transparency are the major features of the blockchain technology, extracting information from the transaction records is an intuitive way to detect phishing scams on the Ethereum platform [9]–[11]. The Ethereum transaction history can be modeled as a directed transaction network, where a node is a unique *address* (We use "address" and "account" interchangeably in this article.) and an edge refers to the existence of at least one transfer of ether between two addresses. Yet when utilizing the transaction records of Ethereum for fraud identification, we may face the following three problems that hinders the performance of fraud identification.

*Extreme data imbalance* is one of the biggest obstacles for phishing detection on Ethereum. According to a report on *etherscan.io*, a famous block explore and analytics platform for Ethereum, the total number of addresses and the total number of transactions of Ethereum are more than 500 million and 3.8 billion, respectively. In contrast, the total number of labeled phishing addresses posted on etherscan.io is only 2041. Therefore, finding phishing addresses in such a huge network is tantamount to finding a needle in a haystack.

*Network heterogeneity* of the Ethereum transaction network refers to the fact that many transactions are related to some public or popular addresses, such as wallet, exchanges, and famous ICOs while the majority of addresses including both normal and phishing addresses may have a relatively small number of transactions. In such a heterogeneous network, it may be more difficult to classify the phishing and nonphishing addresses with topological information only.

*Feature Extraction:* The identification of phishing addresses on Ethereum is essentially a classification problem in machine learning, whose performance is closely related to the choice of data representation and extracted features. Only when we extract the characteristics which can accurately distinguish phishing and nonphishing addresses, can we effectively implement the detection scheme for phishing scams.

Intuitively, the problem of phishing detection on Ethereum can be modeled as a binary classification problem and solved by using supervised learning approaches. However, the aforementioned problems of extreme data imbalance and network heterogeneity may influence the performance of supervised classification methods in a large scale. Therefore, here, we employ an unsupervised anomaly detection approach, namely, one-class support vector machine (SVM), to solve the phishing detection task on Ethereum by turning it into a single classification task.

On the other hand, the performance of machine learning methods is heavily dependent on the choice of data representation (or features). Network embedding is a learning paradigm which embeds nodes, links, or entire (sub) graphs into a low-dimensional. Compared with the traditional feature engineering method, network embedding is a more efficient and automatic way to extract features from large-scale networked data, thus being a promising candidate for the issue of phishing detection on Ethereum discussed in this work. Different from general networks, each link in the Ethereum transaction network has specific transaction information, such as transaction amount and timestamp.

In summary, considering the above-mentioned three challenges, we propose a comprehensive identification model for the detection of phishing scams on Ethereum. First, combining the transaction data obtained through an Ethereum client and the labeled phishing addresses from two authoritative websites, we build a large-scale Ethereum transaction network where the nodes are classified into labeled phishing and unlabeled addresses, and the edges present the transaction between the addresses. Second, to extract features from the large-scale Ethereum transaction network more accurately and efficiently, we design a novel network embedding algorithm called *trans2vec* with biases of transaction amount and timestamp. Finally, to deal with the problems of extreme data imbalance and network heterogeneity, we adopt the one-class SVM to classify the phishing and nonphishing addresses.

The main contributions of this article are summarized as follows.

1) *Problem:* To the best of our knowledge, this work is the *first* investigation on phishing identification on Ethereum via network embedding. The proposed identification model can be utilized by uniform platforms of Ethereum to detect suspicious phishing addresses and remind users when they attempt to transfer money to these suspicious addresses.
2) *Algorithm:* We propose a novel network embedding model specifically for transaction networks by incorporating the transaction amount values and timestamps of transaction links. It is worth mentioning that although the model in this work is proposed for the scenario of phishing identification on Ethereum, it can be applied to behavior recognition scenarios of other similar transaction networks.
3) *Evaluation:* Extensive experiments on an Ethereum transaction network validate the effectiveness of the proposed algorithm in the identification of phishing nodes. Additionally, experimental results demonstrate that the proposed *trans2vec* significantly outperforms the state-of-the-art network embedding methods.

The remainder of this article is organized as follows. Section II presents recent work about frauds on Blockchain, phishing scams, and anomaly detection based on network embedding. In Section III, we provide an overview of the Ethereum data and then give the problem definition of phishing identification on Ethereum. Then, we present the technical details of the proposed embedding algorithm *trans2vec* and the overall detection framework in Section IV. In Section V, we evaluate the phishing detection performance of the proposed method on Ethereum and assess the parameter sensitivity and scalability of *trans2vec*. Finally, we draw conclusions and discuss future work in Section VI.

## II. Background and Related Work

### A. Frauds on Blockchain

Blockchain was invented by Satoshi Nakamoto in 2008 to serve as the public transaction ledger of the cryptocurrency bitcoin. The design of bitcoin has inspired various other applications, and the blockchain technology which has the features of openness and decentralization has been widely used by cryptocurrencies.

Ethereum is an open-source blockchain-based platform featuring smart contract functionality. The concept of Ethereum was first proposed in late 2013, and the formal development of the Ethereum software project began in early 2014.

However, with the rapid development of blockchain technology and applications of smart contracts, there have been a growing number of frauds in the name of digital currency trading and technological innovation [12]. Vasek and Moore [13] performed the first empirical study of financial scams on bitcoin and defined four categories of scams: 1) Ponzi schemes; 2) mining scams; 3) scam wallets; and 4) fraudulent exchanges. On the platform of Ethereum, a number of studies have probed into the vulnerability of smart contracts and the security of ICOs [14]–[18]. For example, Atzei *et al.* [14] analyzed the security of Ethereum smart contracts by discussing the major attacks and threats.

In addition to research on smart contracts and ICO applications, a series of studies focusing on scam detection in the Ethereum platform has been conducted. Bartolett *et al.* [19] reported the first comprehensive investigation of the Ethereum Ponzi scheme. Later, Chen *et al.* [20] put forward a method to identify hidden smart Ponzi schemes through data mining and machine learning approaches.

### B. Phishing Scams

Phishing refers to a form of online threat defined as the art of impersonating a website of an honest firm aiming to acquire users' private information, such as usernames, passwords, and social security numbers [21]. Typically, a traditional phishing attack begins by sending an email that seems to be from an authentic organization to victims.

To counter the threat from phishing scams, a number of anti-phishing solutions have been proposed by both industry and academia. For example, Abdelhamid *et al.* [21] investigated phishing detection using a multilabel classifier, Medvet *et al.* [22] presented a novel technique to visually compare a suspected phishing page with a legitimate one, and Zouina and Outtaj [23] presented a detection system for phishing websites using SVM.

In summary, as traditional phishing frauds usually rely on phishing emails and websites to obtain users' sensitive information, most existing detection methods of phishing scams are based on text detection of email and Web content.

Compared with traditional scenarios, phishing scams on Ethereum can be conducted in more diverse ways. This kind of phishing frauds can not only obtain the users sensitive information and money via phishing websites but also swindle money directly by spreading phishing addresses to victims via emails, websites, and online chats. Take the well-known phishing scam on Bee Token ICO [24] as an example, the phishers sent a fake email to would-be buyers before the startup runs its token sale and induced the investors to transfer money to a particular address. Without any phishing website, this phishing scam eventually gathered about $1 million from the investors in only 25 h.

Therefore, traditional phishing detection methods based on websites cannot be directly applied to solve the phishing detection problem on Ethereum, because only a small part of phishing scams are implemented through phishing websites.

However, thanks to the openness and transparency of blockchain, victims of phishing scams can usually find out where their fraudulent funds went and report the suspicious phishing addresses. Besides, all transaction records on Ethereum are publicly accessible, thus making it possible to identify phishing addresses via mining the transaction behavior between addresses.

To this end, considering the unique characteristics of the Ethereum environment, we model the transaction records between addresses as a directed transaction network and propose a network embedding framework of phishing detection by extracting information from the Ethereum transaction records.

Fig. 1 compares the phishing process and phishing detection framework between traditional scenarios and Ethereum. We can see that the Ethereum phishing detection is different from the conventional way in terms of detection objects, adopted data sources, as well as detection methods. First, unlike traditional detection methods aiming to figure out the phishing websites, our detection object here is to detect the Ethereum addresses of the phishers. Therefore, compared with traditional methods, the phishing detection framework discussed in this article has a distinct detection granularity. Second, traditional detection models are mainly based on content and URL information of the websites [23], [25], [26], while the detection framework here utilizes the transaction records between Etherum addresses to distinguish the phishing and nonphishing addresses. Third, in terms of detection methods, most existing detection methods of phishing scams extract features of websites, such as URLs, hyperlinks, sensitive words hinting at the possibility of phishing, and other content-based features based on text detection of email and Web content, while here we model the Ethereum transaction history as a directed transaction network and propose a network embedding method which can automatically learn the features of addresses to distinguish phishing and nonphishing addresses.

### C. Anomaly Detection Based on Network Embedding

With the explosive growth of big data, network-based anomaly detection algorithms have attracted increasing attention from both academia and industry. Network-based anomaly detection methods can be applied to both static and dynamic graphs with tags or attributes, applicable to many areas, such as financial networks, security, healthcare, and so on [27].

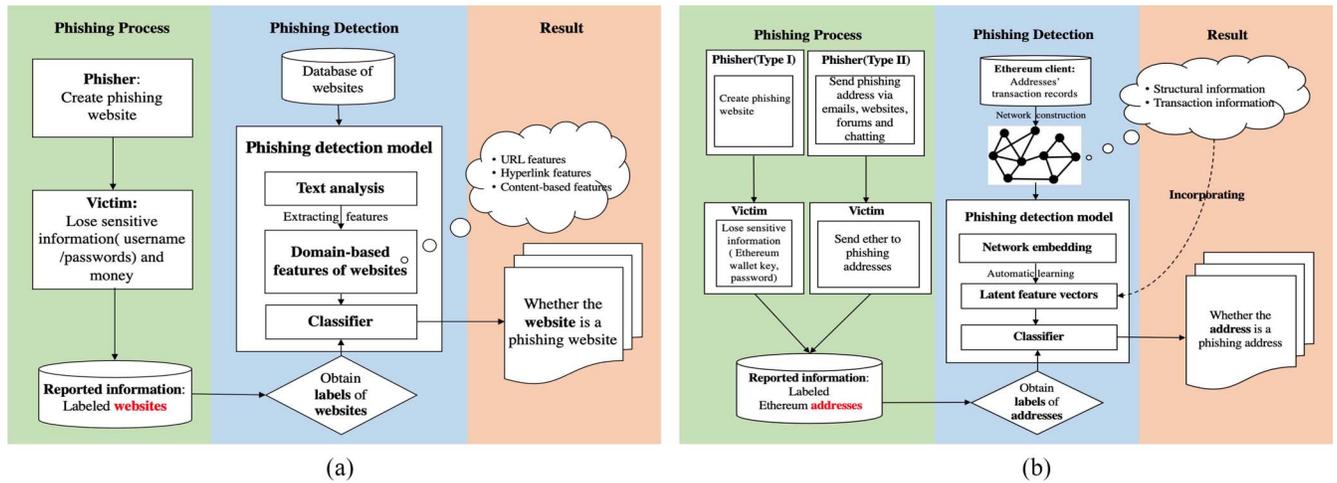

Fig. 1. Comparison of the phishing process and phishing detection framework between (a) traditional scenarios and (b) Ethereum.

Anomaly detection based on network embedding is an emerging technique in recent years. Some of these anomaly detection techniques propose their own embedding methods based on the characteristics of the network [27], [28], while some others adopt existing embedding methods. Existing network embedding methods can be categorized into four main classes: 1) factorization methods [29], [30]; 2) random walk techniques [31], [32]; 3) deep learning [33], [34]; and 4) other miscellaneous strategies [35]. After optimizing the embedding space, the results of the learned node embeddings can be utilized as extracted feature inputs for downstream machine learning tasks.

## III. Data Description and Problem Definition

This study aims to detect the phishing addresses on Ethereum based on large-scale transaction records. In this section, we first provide an overview of the transaction data and then introduce the problem of fault identification on Ethereum.

### A. Data Description

*Transaction Records:* In order to obtain the Ethereum transaction network, we adopt the same method used in [36] to collect transaction records from the launch of Ethereum through an Ethereum client. According to the Ethereum Yellow Paper [37], each Ethereum client contains all transaction history. In this way, we can obtain a transaction record dataset we need, and then build a transaction network where each node represents an address and each edge indicates the ether transaction between a pair of addresses. It should be noted that in this transaction network, each edge between two nodes is assigned with the total transferred amount and the timestamp of the last transaction between them.

*Labeled Phishing Addresses:* The issue of phishing identification on Ethereum can be modeled as a binary classification problem, which is a typical supervised learning task. Therefore, we need enough labeled data to train the classification model and further verify the effectiveness of our method. In this work, we collect the labeled data about phishing scams from two authoritative websites which report various illegal behaviors on Ethereum. One is *EtherScamDB* (https://etherscamdb.info/scams), which collects information about online scams to guide Ethereum investors away from possible frauds. The other one is *Etherscan* (https://etherscan.io/), which serves as an Ethereum block explorer. The reports about various scams on these two websites show not only the content of scams but also the addresses suspected of involvement in frauds. From the various scams reported on these two websites, we extracted addresses which are related to phishing scams. To ensure the accuracy of the labeled data, we crawl all the reports about phishing scams before March 10, 2019, and only the addresses reported by both websites are labeled as phishing addresses.

After obtaining all the transaction records, we obtain more than 500 million addresses and 3.8 billion transaction records. However, only 1259 addresses are labeled as phishing addresses. Therefore, the extreme data imbalance is one of the biggest obstacles for phishing detection on Ethereum. Besides, all addresses on Ethereum can be categorized into several types. As shown in Fig. 2, the red points are labeled as phishing addresses, the blue points are known exchanges, the yellow points are smart contract addresses, and other points are common unknown addresses. The different types of nodes described above tend to behave distinctly in terms of transaction characteristics. For example, some public or popular addresses such as the blue and yellow points may involve in a large number of transactions, while a majority of phishing and common addresses may have a relatively small number of transactions. Therefore, the heterogeneity nature of the Ethereum network may make it more difficult to classify the phishing and nonphishing nodes.

Due to the aforementioned issues of extreme data imbalance and network heterogeneity, it is difficult to obtain decent performance by modeling this detection problem into a supervised binary classification problem. Therefore, here, we adopt an unsupervised anomaly detection approach called one-class SVM by turning this problem into a single classification task [38]. In this way, the behavior of phishing nodes can be distinguishable from the others in a suitable feature space, and the task of detecting the phishing nodes is a "outlier detection" or "one-class classification," which aim to find a decision surface around the *targets*. The nodes that lie inside this decision surface are classified as *targets* (i.e., the phishing nodes), whereas nodes that lie outside are classified as *outliers* (i.e., other nodes).

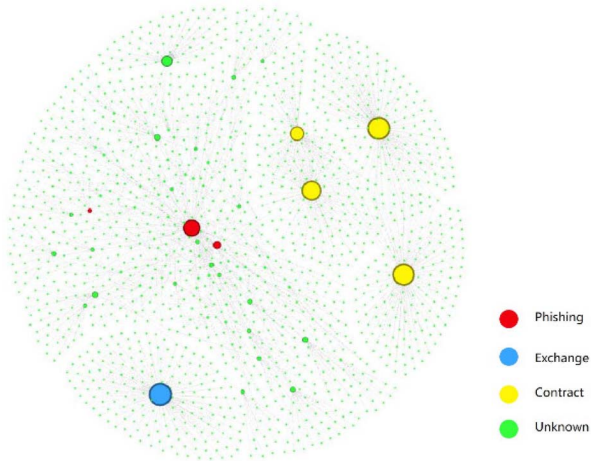

Fig. 2. Simple example of an Ethereum transaction network with four kinds of addresses, namely, phishing addresses, known exchanges, smart contract addresses, and common addresses with unknown labels.

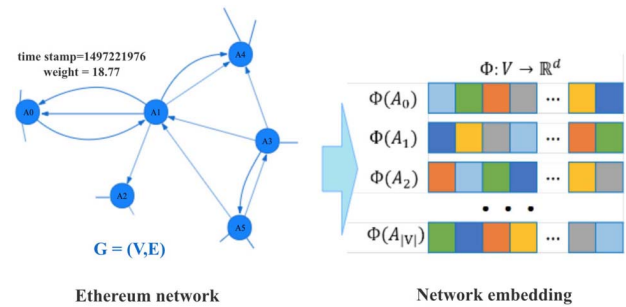

Fig. 3. Process of network construction and embedding for the Ethereum transaction network.

### B. Problem Definition

Through the above operations on the Ethereum data, we obtain a transaction network. Let $G = (V, E)$, where $V$ represent the set of nodes, and $E$ is the set of edges. $G_L = (V, E, X, Y)$ is a partially labeled network, with edge attributes $X \in \mathbb{R}^{|E| \times S}$ where $S$ is the size of the feature space for each edge, and $Y \in \mathbb{R}^{|V| \times |\mathcal{Y}|}$ where $\mathcal{Y}$ is the set of labels. In the Ethereum transaction network, each edge contains two critical attributes, namely, transaction amount and timestamp. For the scenario of phishing address identification, $\mathcal{Y}$ contains two labels, i.e., $+1$ for phishing node and $-1$ for normal samples.

The principal aim of this work is to detect phishing scams on an extremely large-scale Ethereum network. Because of the large-scale of the network and the imbalance of data labels, we propose a biased network embedding algorithm, which incorporates the transaction amount and timestamp of each edge to better capture the information from the Ethereum transaction network. The goal of the network embedding algorithm is to learn the embeddings of all nodes $X_E \in \mathbb{R}^{|V| \times d}$, where $d$ is the number of dimensions for feature representation. These obtained node embeddings can be used as feature inputs for the downstream classification task.

Fig. 3 gives a simple illustration of the embedding procedure on the Ethereum transaction network.

## IV. MODEL FRAMEWORK

In this section, we first introduce the proposed network embedding method for transaction networks called *trans2vec*, including the feature learning process of network embedding and the proposed neighborhood sampling strategy for Ethereum, and then describe the overall framework of phishing detection.

### A. Feature Learning Process

In recent years, random walk-based network embedding has been proposed and widely used to automate the process of feature extraction.

This kind of network embedding aims to learn a mapping function from nodes to node embeddings ($f : V \longmapsto \mathbb{R}^{|V| \times d}$), maximizing the likelihood of co-occurrence of neighbor nodes in a $d$-dimensional feature space. The embedding process consists of two main parts: the first part is a random walk generator, which is used to capture the structural relationships between nodes; and the second part is the Skip-gram architecture, which is used to learn the node embedding via solving a maximum-likelihood optimization problem. By conducting truncated random walks, a large-scale network is transformed into a set of node sequences sampled from it. For each source node $u \in V$, each node sequence sampled from the network by a particular sampling strategy $S$ is defined as $N_S(u) \in V$.

Skip-gram is a widely adopted architecture for data representation learning which was originally proposed for natural language processing [39]. The objective of Skip-gram is maximizing the co-occurrence probability among the words that appear within a window. Inspired by the Skip-gram architecture, network researchers proposed to present a network as a "document."

Following previous studies on network embeddings, here we employ the Skip-gram architecture to optimize the following objective function, which maximizes the log probability of the occurrence of nodes from the neighborhood $N_S(u)$ for a node $u$ conditioned on its node embedding, i.e.,

$$\max_f \sum_{u \in V} \log P_r(N_S(u) \mid f(u)). \tag{1}$$

In this work, we adopt the stochastic gradient descent approach to optimize $f$ by solving (1).

### B. trans2vec for Transaction Networks

As mentioned above, the Skip-gram architecture adopted in random walk-based network embedding methods was originally inspired by the word2vec in natural language processing [40]. As words in natural language are linearly listed, it is reasonable to use sliding windows to define neighbors of words. However, nodes in the network are not linearly connected, and thus we need to use a method to define the neighborhood of a node.

For the Ethereum transaction network, considering the amount value and timestamp of each transaction, we propose three targeted biased random walk strategies for neighbor sampling, which can extract the indispensable information of the transaction network more comprehensively.

*1) Random Walks:* Given a source node $u$, we perform the random walk to obtain node sequences with a fixed length $l$. Let $c_i$ denote the $i$th node in the sequence, starting with $c_0 = u$. The probability of choosing a particular node $x$ as $c_i$ is

$$P(c_i = x \mid c_{i-1} = u) = \begin{cases} \frac{\pi_{ux}}{Z}, & \text{if } (u, x) \in E \\ 0, & \text{otherwise} \end{cases} \quad (2)$$

where $\pi_{ux}$ is the unnormalized transition probability from nodes $u$ to $x$, and $Z$ is the normalizing constant.

*2) Search Strategies:* In a random walk-based network embedding algorithm, it is pivotal to choose proper preferences in the walking process. For example, *node2vec* defines two parameters to interpolate between depth-first search and breadth-first search [32]. This highly adaptable algorithm takes into account both local and global information. In this work, we focus on a type of transaction network which contains some unique information. For financial transaction networks like the Ethereum, each edge has a particular amount and timestamp, which is very critical but cannot be captured by general random walk-based network embedding methods. In order to learn the features of a transaction network more comprehensively, we design biased random walk strategies based on transaction amount and timestamp.

The process of sampling neighborhoods of a node can be viewed as a local search. For the purpose of a fair comparison, we set the size of the neighborhood set as $k$ and then search a number of sets for each node. For the Ethereum transaction network discussed in this work, we consider two kinds of biased sampling methods.

*Amount-Based Biased Sampling:* Intuitively, a larger amount of value of the transaction implies a stronger or closer relationship between the two involved nodes. We denote $V_u$ as the set of nodes directly connected to node $u$ and use a linear function to incorporate the amount information to the sampling probability. Under the amount-based biased sampling, starting from node $u$, the transition probability from node $u$ to a neighbor node $x \in V_u$ is given as

$$PA_{ux} = \frac{A(u, x)}{\sum_{x' \in V_u} A(u, x')} \quad (3)$$

where $A(u, x)$ denotes the total amount value of the transaction(s) between nodes $u$ and $x$.

*Time-Based Biased Sampling:* In addition, each edge has a unique timestamp. Here, we assume that the later the transaction is, the greater the impact on the current relationship of the nodes. First, we map the realistic timestamps of the edges to discrete time steps. Let $T: E \longrightarrow \mathbb{Z}$ be a function that sorts the transaction edges in ascending order of timestamps. Similarly, under the time-based biased sampling, starting from

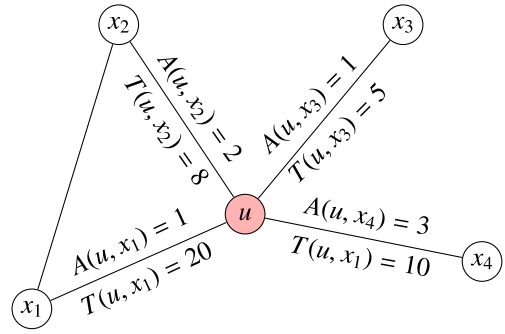

Fig. 4. Illustration of an Ethereum transaction network.

node $u$, the transition probability from node $u$ to a neighbor node $x \in V_u$ is

$$PT_{ux} = \frac{T(u, x)}{\sum_{x' \in V_u} T(u, x')} \quad (4)$$

where $T(u, x)$ denotes the timestamp of the latest transaction between nodes $u$ and $x$.

Fig. 4 plots a simplified illustration of the random walk procedure in a transaction network. Starting from node $u$, the node $x_4$ will be most likely to be chosen as the next node when the amount-based sampling is adopted. While under the time-based biased sampling, node $x_1$ should have the largest probability to be sampled.

*Search Bias Parameter $\alpha$:* In order to take both time and amount into account, we use a parameter $\alpha(0 \leq \alpha \leq 1)$ to balance their effects. The unnormalized transition probability from nodes $u$ to $x$ can be given as

$$\pi_{ux}(\alpha) = PA_{ux}^{\alpha} \cdot PT_{ux}^{1-\alpha}. \quad (5)$$

Here, the parameter $\alpha$ allows the sampling procedure to adjust its bias between time and amount. As shown in Fig. 4, the edge $(u, x_1)$ has a larger value of time step but a smaller value of amount when compared with the edge $(u, x_4)$. The strategy will be more likely to sample node $x_1$ when $\alpha$ is very small and tends to sample node $x_4$ otherwise. In this way, the search bias can be balanced between time and amount weights.

*3) trans2vec Algorithm:* The proposed random walk-based network embedding method is named *trans2vec* as its principal task is to embed the transaction information into node representation vectors. The pseudocode for the proposed *trans2vec* is listed Algorithm 1. We conduct the process of *trans2vec* random walk to sample the large-scale transaction network. Specifically, we perform $r$ random walks with walk length $l$ from each source node. At every step of the walk, we design a biased sampling strategy, in which a search bias parameter $\alpha$ allows us to smoothly transfer between the two biases based on transaction amount and time. It should be noted that as the transition probabilities $\pi_{vx}$ can be precomputed, the random walk procedure of the *trans2vec* can be conducted efficiently in $O(1)$ time using alias sampling. In addition, similar to previous work [32], we first use a preprocessing procedure to calculate the transition probabilities, then conduct the *trans2vec* random walks, and finally, optimize the mapping function $f$ of the node embeddings by utilizing stochastic gradient descent.

**Algorithm 1** trans2vec Algorithm

**Require:** (The transaction network $G = (V, E, X)$ where $X$ contains the transaction amount and timestamp information of all edges, embedding dimension $d$, walks per node $r$, walk length $l$, context/neighborhood size $k$, search bias parameter $\alpha$)
$\pi$ =PreprocessTransitionProbability(G,$\alpha$)
$G' = (V, E, X, \pi)$
Initialize *walks* to Empty
**for** *iter* = 1 **to** $r$ **do**
  **for** *each node* $u \in V$ **do**
    *walk*=trans2vecwalk($G'$, $u$, $l$)
    Append *walk* to *walks*
  **end for**
**end for**
$f$ = StochasticGradientDescent($k$, $d$, *walks*)
**return** $f$

---

**trans2vecwalk** (Graph $G' = (V, E, \pi)$, Starting node $u$, Length $l$, search bias $\alpha$)
Initialize *walk* to [$u$]
**for** *walk_iter* = 1 **to** $l$ **do**
  *curr* = *walk*[-1]
  V_*curr* = GetNeighbors(*curr*, $G'$, $\alpha$)
  $s$ = AliasSample($V_{curr}$, $\pi$)
  Append $s$ to *walk*
**end for**
**return** *walk*

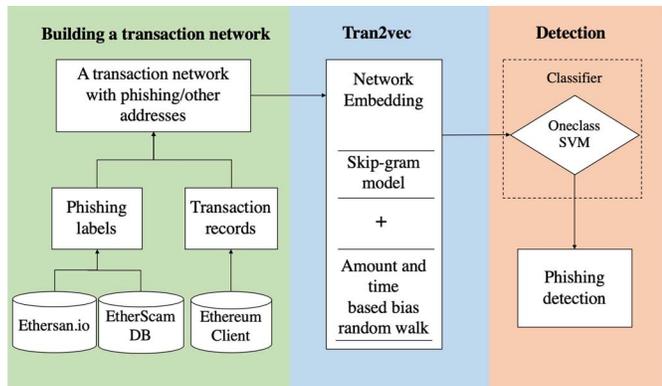

Fig. 5. Framework of phishing detection on Ethererum.

### C. Phishing Detection Framework

After utilizing the proposed *trans2vec* algorithm to obtain node embeddings, we use them as feature inputs for the task of phishing scam detection on Ethereum. Fig. 5 presents the overall framework of the phishing detection method which contains three main steps.

First, combining the collected transaction records through an Ethereum client and the labeled phishing addresses from two authoritative websites, we build a large-scale Ethereum transaction network where the nodes are classified into phishing and other addresses, and the edges present the transaction between each pair of addresses. Second, to extract features from the Ethereum transaction network more accurately and efficiently, we design a novel network embedding algorithm *trans2vec* with biases of transaction amount and timestamp. Finally, we adopt the one-class SVM to classify the phishing and other addresses.

TABLE I
NETWORK PROPERTIES OF THE 50 EXTRACTED SUBNETWORKS

| Network Properties | Node Number | Edge Number | Average Degree |
|---|---|---|---|
| Average | 60442.3 | 236221.6 | 6.62 |
| Maximum | 97221 | 310244 | 7.26 |
| Minimum | 49862 | 162766 | 6.12 |

## V. EXPERIMENTS

In this section, we present the experimental results of the proposed phishing detection framework. First, we describe the experimental dataset. Then, we explain the setup of the experiments in detail. Finally, we perform the phishing detection task on Ethereum transaction data to demonstrate the effectiveness, parameter sensitivity, and scalability of the proposed algorithm and give an analysis of experimental results.

### A. Datasets

As discussed in Section III-A, here we adopt the one-class SVM method to classify phishing and nonphishing nodes. As 1259 addresses are labeled as phishing nodes which are the targets of the detection approach, we randomly select 1259 unlabeled nodes as the outliers.

With these labeled and unlabeled nodes being the central nodes, we extract their first-order neighbors and the connected edges between all of them to form a subnetwork. In our experiments, we repeat the random selection procedure of unlabeled nodes for 50 times and thus obtain 50 subnetworks. As shown in Table I, these subnetworks contains more than 60 000 nodes and 200 000 links on average. Then, each subnetwork is embedded via the proposed *trans2vec* method to obtain the feature vectors of the central nodes for the downstream classification task. In the final classification task, we set 80% of the total data as training data and the rest as test data.

### B. Baseline Methods

Referring to Fig. 5, we propose a biased random walk-based network embedding method to obtain the node feature vector. In the experiment, our proposed method is compared with two popular network embedding approaches based on the random walk, i.e., DeepWalk and node2vec.

1) *DeepWalk [31]:* This is the pioneering work to learn node representations via simulating unbiased random walks. It proposes to sample the network via random walks on the network and defines the neighborhood/context of a node by its co-occurred nodes on the walks. After the process of node sampling, it learns node embeddings by predicting each node's neighborhood.
2) *node2vec [32]:* Following DeepWalk, node2vec defines a more flexible notion of a node's neighborhood and exploits a biased random walk to encode both local and global network structure.

Considering the properties of a transaction network, the proposed *trans2vec* strategy samples the network based on both two kinds of transaction features, including the transaction amount and timestamp. In order to observe the effect of each feature more clearly, we also consider a biased sampling method based on only time or amount in the experiments.

To implement these network embedding methods, we need to set the following parameters: embedding size $d$, walks per node $r$, walk length $l$, and context size $k$. In our experiment, the parameter settings are $d = 64$, $r = 20$, $l = 5$, and $k = 10$. For node2vec, we set $p = 0.25$ and $q = 0.75$ according to the guidance given in [32]. For the proposed *trans2vec*, we vary the search bias parameter $\alpha$ from 0 to 1, and set $\alpha = 0.5$ as the default value to balance the effects of amount and time biases.

However, all the aforementioned network embedding methods learn and encode the topological structural information of the Ethereum transaction network automatically. Therefore, in order to verify the importance of the network structural information as well as the time and amount information of the transactions, we consider three nonembedding methods to extract local features of the addresses for phishing detection, namely, time features only method, amount features only method, as well as time plus amount features method. In detail, the time features only method extracts the maximum time interval, the minimum time interval, total transaction time, and trading frequency of each address; the amount features only method extracts the maximum transaction amount, the minimum transaction amount, total transaction amount, and average transaction amount of each address; and the time plus amount features method consider the above eight statistics as extracted features.

### C. Performance Evaluation Metrics

To evaluate the performance of different methods in terms of phishing detection, we consider three evaluation metrics, namely, precision, recall, and F-score. We repeat experiments on each subnetwork for 100 times and report the average results.

The three metrics are defined as follows:

$$\text{Precision} = \frac{\text{true positive}}{\text{true positive} + \text{false positive}}$$
$$\text{Recall} = \frac{\text{true positive}}{\text{true positive} + \text{false negative}}$$
$$\text{F-score} = 2 \times \frac{\text{Precision} \times \text{Recall}}{\text{Precision} + \text{Recall}}.$$

### D. Classification Performance

We first compare the results of the nonembedding methods which have not consider the structural information, and the experimental results given in Table II indicate that all these methods cannot achieve the satisfying performance of phishing detection. Moreover, we can observe that considering the combination of the two kinds of features leads to better classification performance than using only time or amount feature. This result indicates that without structural information, only time or/and amount features of the addresses are not sufficient to achieve decent classification performance.

TABLE II
PERFORMANCE COMPARISONS OF NONEMBEDDING ALGORITHMS WHICH CONSIDER ONLY TIME FEATURES, AMOUNT FEATURES, AND BOTH. THE BEST RESULTS ARE MARKED IN BOLD

| Method | Precision | Recall | F-score |
|---|---|---|---|
| Time Features Only | 0.351 | 0.302 | 0.326 |
| Amount Features Only | 0.396 | 0.321 | 0.358 |
| Time + Amount Features | **0.509** | **0.478** | **0.494** |

TABLE III
PERFORMANCE COMPARISONS OF DIFFERENT EMBEDDING ALGORITHMS WHEN THE EMBEDDING DIMENSION $d$ IS SET AS 64. THE BEST RESULTS ARE MARKED IN BOLD

| Method | Precision | Recall | F-score |
|---|---|---|---|
| Deepwalk | 0.799 | 0.762 | 0.780 |
| Node2vec | 0.870 | 0.822 | 0.845 |
| Time-based Bias | 0.864 | 0.822 | 0.842 |
| Amount-based Bias | 0.883 | 0.855 | 0.868 |
| **trans2vec** | **0.927** | **0.893** | **0.908** |

Given embedding size $d = 64$, the experimental results of the embedding methods which encode network structural information are compared in Table III. The results given in Table III demonstrate that the proposed *trans2vec* method outperforms the other embedding methods in terms of all evaluation metrics. Moreover, we can observe that both the amount-based and time-based samplings perform better than the unbiased DeepWalk, and a comparison between these two biases indicates that the amount factor tends to have a more important influence on the embedding results for the Ethereum transaction network, and thus achieve better performance than the time-based bias. Therefore, Table III indicates that extracting only structural information cannot ensure good performance. After incorporating the time and amount features of the transactions with structural information, the proposed embedding method performs best.

Combining the results in Tables II and III, we can conclude that the structural information of the transaction networks, as well as the transaction time and amount are indispensable features for the phishing detection task.

As shown in Fig. 6, we gradually increase the embedding dimension of the node vectors from 4 to 64. Obviously, the larger the node vector dimensions, the better the classification performance. This is because larger node vector dimensions are likely to retain richer network structure and node information. We observe that when the embedding dimension $d$ is set as 4, the performance of *trans2vec* in terms of recall and F-score is relatively worse than that of the amount-based biased method, implying that the *trans2vec* method can embed richer information and thus requires a larger value of $d$ for data representation. When we select a larger value of the embedding dimension ($d \geq 8$ in Fig. 6), *trans2vec* performs best in terms of precision, recall, and F-measure.

Besides, the selected classifier of the detection framework is also a factor affecting the detection performance. Therefore, here we consider several widely considered classifiers as baselines, namely, logistic regression, naive Bayes, and isolation forest. Using the node representation vectors of trans2vec with dimension $d = 64$ as input features, the detection results of different classifiers are compared in Table IV. We can observe that the performance of the one-class SVM is obviously better than other classifiers as it is more suitable for the problem of anomaly detection here, and thus we select it as the classifier in our phishing detection framework.

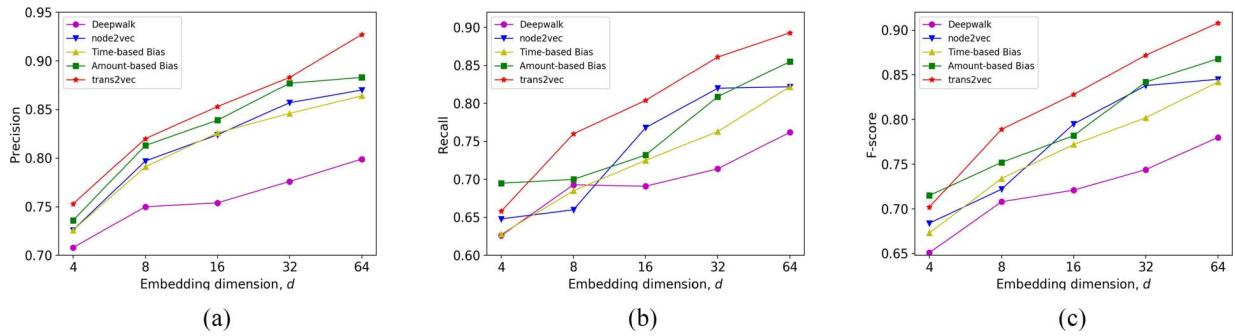

Fig. 6. Performance comparisons of different methods with various embedding dimensions $d$ in terms of (a) precision, (b) recall, and (c) F-score.

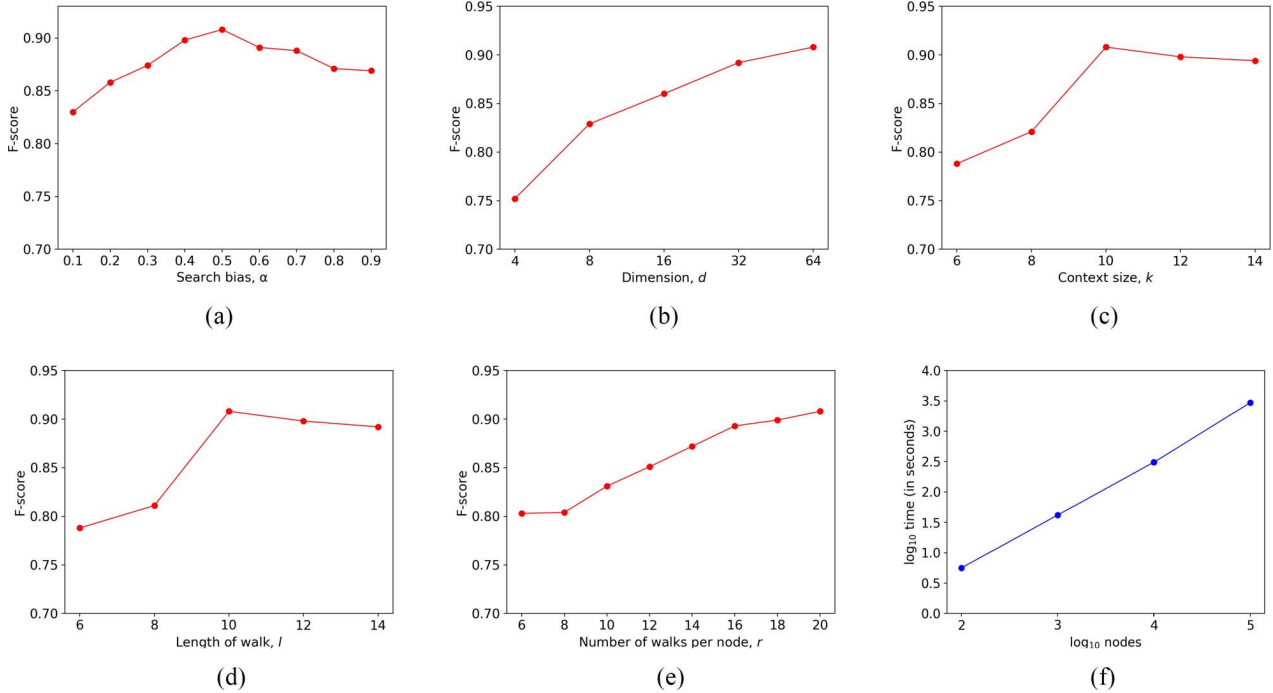

Fig. 7. Results of (a)–(e) parameters analysis and (f) scalability.

TABLE IV
PERFORMANCE COMPARISONS OF DIFFERENT CLASSIFIERS WHEN THE FEATURES ARE EXTRACTED USING THE PROPOSED TRAN2VEC WITH $d = 64$. THE BEST RESULTS ARE MARKED IN BOLD

| Method | Precision | Recall | F-score |
| --- | --- | --- | --- |
| Logistic regression | 0.762 | 0.738 | 0.75 |
| Naive bayes | 0.771 | 0.702 | 0.736 |
| Isolation forest | 0.821 | 0.849 | 0.835 |
| One-class SVM | **0.927** | **0.893** | **0.908** |

### E. Parameter Sensitivity

For the proposed *trans2vec*, there exists a number of parameters which may influence the embedding results. In Fig. 7(a)–(e), we evaluate the effects of a series of parameters on the performance of *trans2vec* on the phishing detection task on the Ethereum transaction network. When a particular parameter is under evaluating, all other parameters are set as default values. In this part, we only consider F-measure for performance comparison.

We first explore the effect of $\alpha$ on F-measure by varying $\alpha$ from 0.1 to 0.9. As shown in Fig. 7(a), the peak value appears when $\alpha$ is around 0.5. This result indicates that the combination of these two biases can achieve better classification performance. When $\alpha$ is set as 0, the algorithm becomes time-based bias sampling. While when $\alpha$ is set to 1, the algorithm becomes pure amount-based bias sampling.

We also examine the influence of the embedding dimension $d$ and the node's neighborhood parameters, including the number of walks $r$, walk length $l$, and neighborhood size $k$. As shown in Fig. 7(b), with an increase of the embedding dimension $d$, the algorithm can achieve better detection performance. Besides, we observe from Fig. 7(c) that an increase of the context size $k$ from 6 to 10 can improve F-measure obviously but the performance seems to saturate when $k$ reaches 10. Similarly, referring to Fig. 7(d), increasing the length of walk $l$ from 2 to 6 can boost the performance. However, when $l$ continues to increase, the algorithm will always walk to the same node, thus reducing the quality of node representations and the overall performance. Fig. 7(e) indicates that a larger number of walks per node also improve the performance, which is not surprising because it indicates a larger number of sampling times to learn network representations.

*F. Scalability*

To evaluate the scalability of *trans2vec*, we conduct this algorithm with default parameter values for Erdos–Renyi (ER) random graphs with node sizes increasing from $10^2$ to $10^5$. For each network size, we do 100 independent trials and compute the average running time. As the ER random graphs are generated using a theoretical complex network model, the edges cannot contain the transaction amount or timestamp which is required by the *trans2vec* to calculate the transition probability. To this end, we set the transaction amount and timestamp of each edge in the random graphs as 1 to facilitate a similar calculation.

The results of running time (in log scale) are shown in Fig. 7(f). We observe that *trans2vec* scales linearly with the number of nodes, which is acceptable in practice. Therefore, we can conclude that *trans2vec* is a scalable method which is suitable for applications on large-scale networks.

## VI. CONCLUSION AND FUTURE WORK

In this article, we conducted the *first* systematic study of phishing scams detection on Ethereum via network embedding. Specifically, a three-step framework was proposed to identify phishing nodes using their features extracted from Ethereum transaction history with network embedding algorithms. To extract features from the Ethereum transaction network more accurately and efficiently, we designed a novel network embedding algorithm *trans2vec* with biases of transaction amount and timestamp. Experiments on real-world Ethereum transaction records demonstrated the effectiveness of our proposed detection framework and the superiority of *trans2vec* over baseline methods in terms of feature extraction for Ethereum-like transaction networks.

Though urgent and important, the problem of phishing scam detection on Ethereum is still unexplored till now. As a preliminary work in this area, we hope this work can attract extensive attention and efforts in this field. Several important research issues can be explored along this topic. First, with more comprehensive domain knowledge and a more detailed data analysis, a more systematic and generalized network embedding algorithm can be proposed for Ethereum and other large-scale transaction networks. Second, as this article focuses on the problem of phishing detection, effects of the proposed network embedding on other realistic downstream tasks remain to be verified. Third, detection and prevention methods for other illegal behaviors on Ethereum, such as gambling, money laundry, Ponzi schemes, etc., can be proposed by utilizing the openness nature of the blockchain technology.

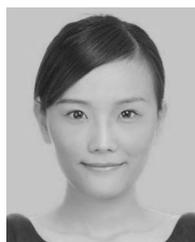

**Jiajing Wu** (Senior Member, IEEE) received the B.Eng. degree in communication engineering from Beijing Jiaotong University, Beijing, China, in 2010, and the Ph.D. degree in electronic and information engineering from Hong Kong Polytechnic University, Hong Kong, in 2014.

In 2015, she joined Sun Yat-sen University, Guangzhou, China, where she is currently an Associate Professor. Her research focus includes blockchain, graph mining, and network science.

Dr. Wu was a recipient of the Hong Kong Ph.D. Fellowship Scheme during her Ph.D. study in Hong Kong from 2010 to 2014. She serves as an Associate Editor for the IEEE TRANSACTIONS ON CIRCUITS AND SYSTEMS—PART II: EXPRESS BRIEFS.

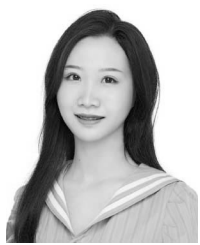

**Qi Yuan** received the B.Eng. degree in computer science and technology from Anhui University, Hefei, China, in 2018. She is currently pursuing the M.Sc. degree in computer science and technology with the School of Data and Computer Science, Sun Yat-sen University, Guangzhou, China.

Her current research interests include fraud detection on blockchain, and theories and applications of network science.

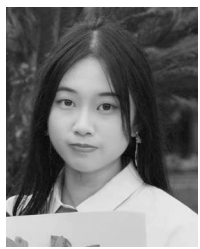

**Dan Lin** received the B.Eng. degree in software engineering from Sun Yat-sen University, Guangzhou, China, in 2019, where she is currently pursuing the M.Sc. degree in computer science and technology with the School of Data and Computer Science.

Her current research interests include blockchain, theories and applications of network science, and machine learning with graphs.

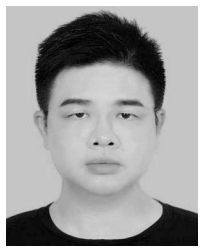

**Wei You** received the B.Eng. and M.Sc. degrees in computer science and engineering from Sun Yat-sen University, Guangzhou, China, in 2017 and 2019, respectively.

His current research interests include graph mining and recommendation systems.

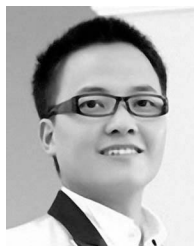

**Weili Chen** received the Ph.D. degree in computer science and technology from the School of Data and Computer Science, Sun Yat-sen University, Guangzhou, China, in 2019.

He is currently a Postdoctoral Associate Research Fellow with Sun Yat-sen University. His research interests include blockchain, data mining, and machine learning.

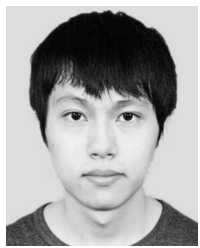

**Chuan Chen** (Member, IEEE) received the B.S. degree in mathematics from Sun Yat-sen University, Guangzhou, China, in 2012, and the Ph.D. degree in mathematics from Hong Kong Baptist University, Hong Kong, in 2016.

He is currently a Research Associate Professor with the School of Data and Computer Science, Sun Yat-sen University. His current research interests include machine learning, numerical linear algebra, and numerical optimization.

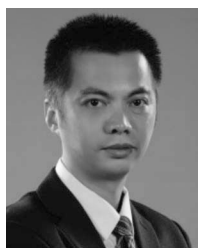

**Zibin Zheng** (Senior Member, IEEE) received the Ph.D. degree in computer science and engineering from the Chinese University of Hong Kong, Hong Kong, in 2011.

He is currently a Professor of Data and Computer Science with Sun Yat-sen University, Guangzhou, China. He serves as the Chair of the Software Engineering Department, Pearl River Young Scholars, and the Founding Chair of the Services Society Young Scientists Forum. In the past five years, he published over 120 international journal and conference papers, including three ESI highly cited papers and 40 ACM/IEEE TRANSACTIONS papers. According to Google Scholar, his papers have more than 6300 citations, with an H-index of 41. His research interests include blockchain, services computing, software engineering, and financial big data.

Dr. Zheng was a recipient of several awards, including the Outstanding Thesis Award of CUHK in 2012, the ACMSIGSOFT Distinguished Paper Award at ICSE2010, the Best Student Paper Award at ICWS2010, and the IBM Ph.D. Fellowship Award. He served as the CollaborateCom'16 General Co-Chair, the ICIOT'18 PC Co-Chair, and the IoV'14 PC Co-Chair.